\documentclass{iau}

\usepackage{amsmath}
\usepackage{graphicx}
\usepackage{multirow}
\usepackage{natbib} 

\begin{document}

\lefttitle{Colzi et al.}
\righttitle{Isotopic ratios in the Galactic Center}

\jnlPage{1}{7}
\jnlDoiYr{2026}
\doival{10.1017/xxxxx}

\aopheadtitle{Proceedings IAU Symposium}
\editors{M. Zaja\v{c}ek,  T. Je\v{r}\'{a}bkov\'{a}, V. Karas, R. Schödel \&  P. Sukov\'{a}, eds.}

\title{Isotopic ratios as probes of star formation and the origin of molecular gas in the Central Molecular Zone}

\author{Laura Colzi}
\affiliation{Centro de Astrobiología (CAB), CSIC-INTA, Ctra. de Ajalvir Km. 4, 28850, Torrejón de Ardoz, Madrid, Spain \\ 
email: {\tt lcolzi@cab.inta-csic.es}}

\begin{abstract}
The Central Molecular Zone (CMZ) contains most of the dense molecular gas in the Milky Way but forms stars at a surprisingly low rate. Isotopic ratios provide a unique tool to investigate both the local physical conditions and the nucleosynthetic history of this gas. I summarize recent isotopic studies toward the molecular cloud G$+$0.693$-$0.027. Deuterium fractionation (D/H ratios) reveals a cold ($T_{\rm kin}\lesssim30$ K), relatively quiescent gas component embedded within the otherwise warm and turbulent CMZ, indicating that the earliest stages of star formation can survive in this extreme environment. I also present new $^{12}$C/$^{13}$C measurements from HC$_3$N and HC$_5$N, including the first constraints from doubly substituted $^{13}$C isotopologues. After accounting for isotopic fractionation with astrochemical models, the inferred elemental ratio ($\sim37$--48) exceeds the canonical CMZ assumed value ($\sim$20), supporting a scenario in which the Galactic center is replenished by less chemically processed gas flowing inward along the Galactic bar.

\end{abstract}


\maketitle
\vspace{-0.3em}
\section{Introduction}

The Central Molecular Zone (CMZ), corresponding to the inner $\sim500$ pc of the Milky Way, is one of the most extreme environments in our Galaxy. It contains nearly 80\% of the dense molecular gas in the Galactic disk, yet its current star formation rate is approximately one order of magnitude lower than expected from standard star formation relations \citep{Longmore2013,Barnes2017}. Understanding the origin, physical properties, and chemical evolution of the molecular gas in the CMZ is therefore essential to explain why star formation proceeds so differently from that in the Galactic disk.

The molecular clouds in the Galactic center are characterized by high kinetic temperatures ($T_{\rm kin}\sim50-150$ K), dust temperature of $\sim$20 K, large velocity dispersions and enhanced turbulence (typically $15-50$ km\,s$^{-1}$), elevated cosmic-ray ionization rates ($\sim$10$^{-15}$ s$^{-1}$), and widespread shocks (\citealt{henshaw2016,henshaw2023,zeng2020,sanz-novo2024,battersby2025}). These extreme conditions significantly modify both the chemistry and dynamics of the gas and have long been proposed as possible mechanisms responsible for suppressing star formation (\citealt{kruijssen2014,henshaw2023}). Nevertheless, star formation is not completely inhibited, as demonstrated by active regions such as Sgr~B2, Sgr~C, the Arches and Quintuplet clusters, and more recently by the detection of embedded dense cores in CMZ clouds (\citealt{portegieszwart2010,nogueras-lara2020,lu2019a,colzi2024}).

\begin{figure}[t]
    \centering
    \includegraphics[width=\textwidth]{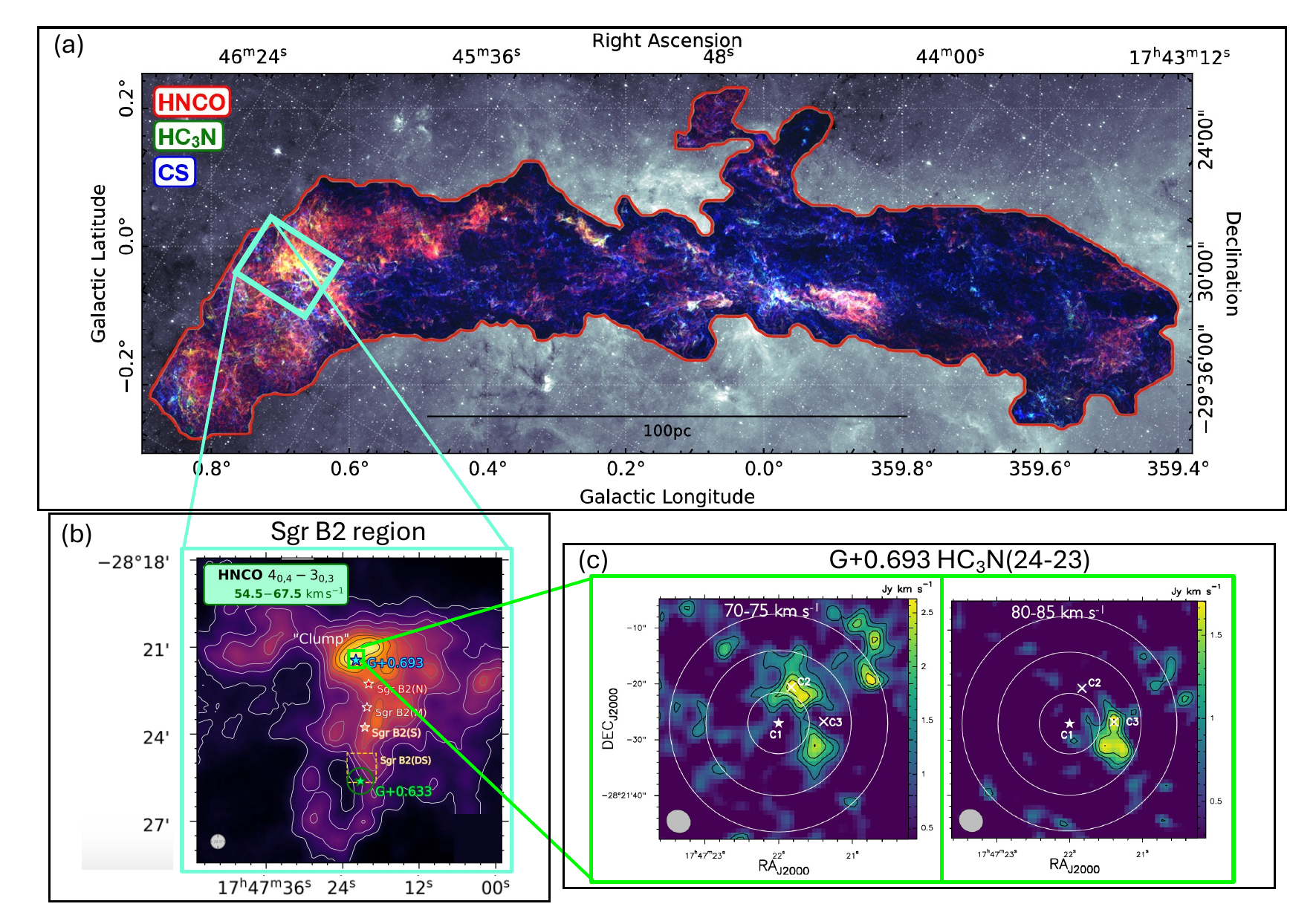}
    \caption{Overview of the Central Molecular Zone (CMZ), highlighting the location of the molecular cloud G$+$0.693$-$0.027 with respect to Sgr~B2. Panel (a) shows the molecular overview of the CMZ from the ACES ALMA survey, adapted from Fig.~3 of \citet{longmore2026}. Panel (b) presents the HNCO($4_{0,4}$--$3_{0,3}$) emission toward the Sgr~B2 region observed with the IRAM 30m telescope, adapted from Fig.~5 of \citet{sanandres2026}. Panel (c) shows G+0.693 as traced by HC$_3$N($24$--$23$) emission observed with the SMA and APEX, adapted from Fig.~5 of \citet{colzi2024}. The labels C2 and C3 identify candidate compact molecular regions that are potential sites of the next generation of star formation in the CMZ.}
    \label{fig:CMZ}
\end{figure}

One of the best laboratories to investigate the chemistry of the Galactic center is the molecular cloud G$+$0.693$-$0.027 (hereafter G+0.693), located about $55''$ northeast of Sgr~B2(N) (Fig.~\ref{fig:CMZ}). Although no active star formation has been identified toward this cloud, it exhibits one of the richest molecular inventories known in the interstellar medium, likely as a result of shocks generated by cloud--cloud collisions (\citealt{zeng2020,colzi2024}). Over the past few years, 30 new interstellar molecules have been discovered toward G$+$0.693, including several prebiotic species containing the key chemical elements essential for life \citep{rivilla2019,rivilla2020,rodriguez-almeida2021a,jimenez-serra2022,rivilla2022b,rivilla2023,sanz-novo2024,sanz-novo2025,sanandres2024,rey-montejo2024,zeng2025,araki2026,jimenez-serra2026}. The remarkable chemical richness of this source makes it an excellent target to investigate the interplay between physical conditions, astrochemistry, and Galactic chemical evolution.

Isotopic ratios provide one of the most powerful tools to study both the local physical conditions and the evolutionary history of the interstellar medium (\citealt{wilson1994,kobayashi2011,romano2022}). On one hand, isotopic fractionation depends on chemical reactions that are strongly affected by temperature, density, the presence of a radiation field or cosmic-ray ionization, allowing isotopic abundances to probe the physical conditions within molecular clouds (\citealt{visser2009,roueff2007,roueff2015,loison2019,colzi2020,sipila2023}). On the other hand, elemental isotopic ratios are determined by stellar nucleosynthesis and Galactic chemical evolution, carrying information on the origin and processing history of the gas (\citealt{romano2017,kobayashi2020,colzi2022b}). Disentangling these two effects is therefore essential to correctly interpret isotopic measurements (e.g. \citealt{colzi2026}).

In this contribution, I summarize recent observational results obtained toward the Galactic center using isotopic ratios. First, I discuss how deuterium fractionation (molecular D/H ratios) can reveal cold, quiescent gas potentially associated with the earliest phases of star formation despite the harsh environment of the CMZ. I then present recent measurements of the carbon isotopic ratio ($^{12}$C/$^{13}$C ratios) based on multiple transitions of HC$_3$N and HC$_5$N, including the first constraints obtained from doubly substituted $^{13}$C isotopologues. Finally, I discuss the implications of these measurements for the origin of the molecular gas in the Galactic center and for ongoing Galactic chemical evolution models that explicitly include isotopes.
\vspace{-1em}

\section{Deuterium fractionation: unveiling the cold gas hidden in the CMZ}

One of the major challenges in studying the CMZ is identifying the earliest stages of star formation. The large molecular column densities, high kinetic temperatures, and broad linewidths characteristic of Galactic center clouds make it difficult to isolate the signatures of cold and dense prestellar material from the surrounding warm turbulent gas. Since the physical conditions in the CMZ differ substantially from those in nearby molecular clouds, alternative tracers are required to identify cold gas that may be embedded within the turbulent medium.

Deuterium fractionation (D/H ratios) provides one of the most powerful diagnostics of the early phases of star formation. At low temperatures ($T_{\rm kin}\lesssim30$ K), isotopic exchange reactions involving H$_3^+$ and its deuterated isotopologues become highly efficient, enhancing the abundance of deuterated molecules by several orders of magnitude above the elemental D/H ratio. As the gas temperature increases, these reactions rapidly become inefficient, and the molecular D/H ratios approach the cosmic abundance. Consequently, enhanced deuterium fractionation is widely used as a tracer of cold, dense, and chemically young gas \citep[e.g.][]{herbst2003,caselliceccarelli2012,roueff2013}.

To investigate whether such cold gas exists in the Galactic center, we measured the D/H ratios of HCN, HNC, HCO$^+$ and N$_2$H$^+$ toward G$+$0.693 using multiple rotational transitions of both the main and deuterated isotopologues \citep{colzi2022a}. The simultaneous analysis of several transitions allowed us to constrain the excitation conditions and revealed that the molecular emission is best reproduced by two distinct velocity components.

The first component exhibits linewidth of approximately $20$ km\,s$^{-1}$ and an excitation temperature of about 7 K. This broad component is representative of the widespread turbulent warm molecular gas commonly observed throughout the CMZ. In contrast, the second component is significantly narrower, with linewidths of only $\sim9$ km\,s$^{-1}$ and lower excitation temperatures of $3-4$ K. The narrow component becomes particularly evident in the higher-$J$ transitions and is the only component detected in DCO$^+$ and N$_2$D$^+$, suggesting that it traces physically distinct material.

The derived deuterium fractions reveal markedly different chemical conditions in the two components. The broad component exhibits D/H ratios in HCO$^+$ and N$_2$H$^+$ that are consistent with the cosmic elemental abundance, while only HCN and HNC show a moderate enhancement. These results are compatible with gas kinetic temperatures above $\sim70$ K, where deuterium fractionation is expected to be largely suppressed (e.g. \citealt{roueff2013}). In contrast, all four molecular families present enhanced D/H ratios in the narrow component, requiring gas temperatures below approximately 30 K. Such low temperatures indicate the presence of cold and relatively quiescent material embedded within the otherwise warm and turbulent molecular cloud.

The existence of this cold component has important implications for our understanding of star formation in the Galactic center. Although the CMZ is often regarded as uniformly warm, our results demonstrate that localized regions with physical conditions similar to those found in nearby prestellar clusters can survive within the extreme Galactic center environment. The narrow component therefore represents a promising candidate for the earliest evolutionary stages preceding star formation. Subsequent excitation and spatial analyses of HC$_3$N confirmed that these narrow linewidth components are associated with denser (2$\times$10$^5$ cm$^{-3}$) localized gas that may eventually collapse and form stars (\citealt{colzi2024} and Fig~\ref{fig:CMZ}).

These results demonstrate that deuterium fractionation remains an effective probe of cold gas even under the extreme environmental conditions of the Galactic center. Combined with complementary molecular tracers, D/H ratios provide a unique tool to disentangle the multiple gas components present along the line of sight and to identify regions that would otherwise remain hidden within the highly turbulent CMZ.
\vspace{-0.5em}
\section{Carbon isotopic ratios and the origin of the molecular gas}

\begin{figure}
\centering
\includegraphics[width=\linewidth]{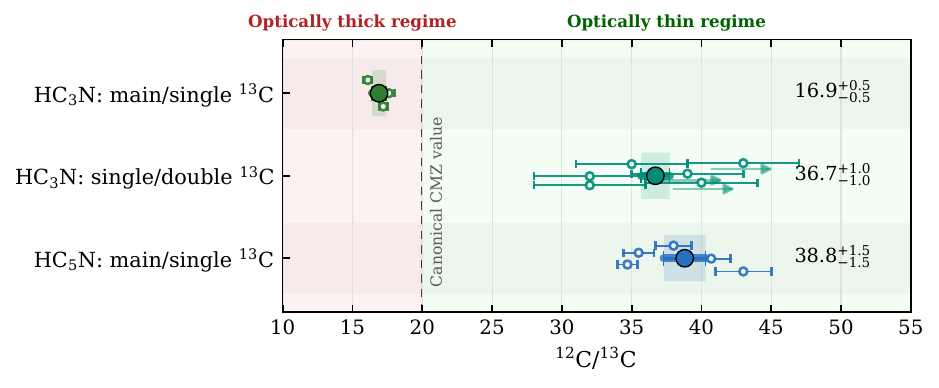}
\caption{Comparison of the $^{12}$C/$^{13}$C ratios derived from different molecular tracers toward G+0.693. Measurements based on the main HC$_{3}$N isotopologues are affected by high optical depth and provide lower limits to the isotopic ratio, whereas analyses of rare single- and doubly substituted $^{13}$C isotopologues yield optically thin estimates of the true carbon isotopic ratio. The agreement between HC$_3$N and HC$_5$N indicates an elemental ratio significantly higher than the canonical value usually assumed for the CMZ. Adapted from \citet{colzi2026}.}
\label{fig:c-isotope-ratio}
\end{figure}

While deuterium fractionation probes the local physical conditions within molecular clouds, isotopic ratios of other elements provide information on much larger spatial and temporal scales. The $^{12}$C/$^{13}$C ratio is primarily determined by stellar nucleosynthesis and Galactic chemical evolution. Since $^{12}$C is predominantly a primary nucleosynthetic product whereas $^{13}$C is mainly produced as a secondary element, the isotopic ratio decreases as the interstellar medium undergoes successive cycles of stellar processing. Consequently, measurements of the carbon isotopic ratio offer a powerful diagnostic of the evolutionary history of molecular gas.

Historically, the Galactic center has been characterized by a canonical $^{12}$C/$^{13}$C ratio of $\sim20-25$, derived from a variety of molecular tracers (e.g. \citealt{fomalont1973,riquelme2010,belloche2013}). However, these measurements are often based on some of the most abundant main isotopologues, whose emission can be affected by optical depth effects, or on single transitions that may not adequately constrain the excitation conditions. Furthermore, chemical isotopic fractionation can modify the molecular isotopic ratios with respect to the underlying elemental abundance, introducing an additional source of uncertainty (\citealt{colzi2020}).

To overcome these limitations, we carried out a comprehensive analysis of multiple rotational transitions of HC$_3$N and HC$_5$N toward G$+$0.693, combining observations of the main isotopologues, the three singly substituted $^{13}$C isotopologues, and, for the first time in the Galactic center, the doubly substituted $^{13}$C isotopologues of HC$_3$N. The simultaneous LTE modelling of numerous transitions spanning a wide range of upper-level energies allows accurate determination of the column densities while minimizing systematic uncertainties associated with excitation effects (\citealt{colzi2026}).

The analysis of the doubly substituted isotopologues provides a direct determination of the carbon isotopic ratio that is largely insensitive to the optical depth affecting the main species. We derive a value of $^{12}\mathrm{C}/^{13}\mathrm{C}=36.7\pm1.0$, which is substantially larger than the value inferred from the main isotopologue alone ($\sim16.9$) and significantly higher than the canonical Galactic center ratio. An independent analysis of HC$_5$N yields a consistent value of $^{12}\mathrm{C}/^{13}\mathrm{C}=38.8\pm1.5$, demonstrating that both cyanopolyynes trace the same isotopic composition. The agreement between these independent molecular families strengthens the robustness of the derived elemental ratio (Fig.~\ref{fig:c-isotope-ratio}).

Because molecular isotopic ratios can be altered by chemical fractionation, the observational results were interpreted using dedicated astrochemical models that explicitly include isotopic chemistry \citep{colzi2020,sipila2023}. Under the physical conditions derived for the warm and more diffuse component of G$+$0.693 ($T_{\rm kin}\simeq140$ K and $n_{\rm H_2}\sim2\times10^4$ cm$^{-3}$, \citealt{colzi2024}), the fractionation degree, $\varphi$, of HC$_3$N and HC$_5$N is found to remain relatively modest, with correction factors of at most $\varphi\sim1.2$. These calculations indicate that the observed molecular ratios correspond to an initial elemental carbon isotopic ratio between approximately 37 and 48, confirming that the high measured values cannot be explained solely by chemical fractionation.

The carbon isotopic ratios measured toward G$+$0.693 are among the highest reported for molecular clouds in the CMZ and are instead comparable to those observed at Galactocentric distances of approximately 4 kpc. This result suggests that at least part of the molecular gas currently present in the Galactic center has experienced a different nucleosynthetic history than expected for material residing permanently in the CMZ. A plausible interpretation is that relatively less processed gas has recently been transported inward through the Galactic bar, replenishing the central molecular reservoir. Hydrodynamical simulations predict that bar-driven gas inflow efficiently channels molecular material from the inner Galactic disk toward the central few hundred parsecs, providing a natural mechanism for producing the observed isotopic signatures (\citealt{tress2020}).

These observational constraints are beginning to be incorporated into new Galactic chemical evolution models for the Nuclear Stellar Disk. Recent models that include gas inflow from the Galactic disk reproduce the observed stellar abundances of Si, Mg, and Ca significantly better than closed-box scenarios, indicating that continuous replenishment of the CMZ by external material is likely to play an important role in shaping its present-day chemical composition (\citealt{ryde2015,sormani2022,spitoni2026}). Future models including different isotopes of elements, together with improved stellar yields and astrochemical corrections, will provide an even more stringent test of this evolutionary scenario.

Overall, the combination of sensitive molecular spectroscopy, observations of multiple isotopologues, and astrochemical modelling demonstrates that carbon isotopic ratios constitute one of the most powerful diagnostics currently available to investigate the origin and evolution of molecular gas in the Galactic center.
\vspace{-0.6em}
\section{Future prospects: extending isotopic studies of the Galactic Center}

The recent results obtained from deuterium fractionation and carbon isotopic ratios demonstrate the diagnostic power of isotopic studies for disentangling the physical and chemical evolution of the molecular gas in the CMZ. Nevertheless, carbon represents only one piece of a much broader picture. Isotopic ratios involving nitrogen, oxygen, sulfur, and other elements provide complementary constraints on stellar nucleosynthesis, gas accretion, and chemical enrichment, allowing the evolutionary history of the Galactic center to be reconstructed in considerably greater detail.

The preliminary analyses of additional isotopic systems toward G$+$0.693 already indicate that the Galactic center may exhibit anomalously high isotopic ratios beyond carbon. Preliminary measurements of the nitrogen isotopic ratio based on HC$_3$N and other molecular tracers suggest significantly large values (e.g. $\sim$1000 from NH$_{3}$ by \citealt{guesten1985a}), while recent analyses of sulfur-bearing molecules indicate enhanced $^{32}$S/$^{34}$S and $^{32}$S/$^{33}$S ratios compared to canonical Galactic center values. Although these studies are still ongoing, they point toward a common scenario in which several isotopic systems reflect the contribution of relatively less chemically processed material to the central molecular reservoir.

Interpreting these observations requires combining molecular spectroscopy with both astrochemical and Galactic chemical evolution (GCE) models. While GCE models predict the elemental isotopic abundances resulting from stellar nucleosynthesis over the lifetime of the Galaxy, the observed molecular isotopic ratios can be modified by chemical fractionation occurring within molecular clouds. Consequently, reliable interpretations require both effects to be considered simultaneously. Recent astrochemical calculations by \citet{colzi2026} show that isotopic fractionation remains relatively modest for the warm conditions prevailing in G$+$0.693, providing confidence that the measured molecular ratios closely trace the underlying elemental abundances. Future work will incorporate increasingly sophisticated chemical networks and models of the physical conditions within CMZ molecular clouds, including shocks, together with updated stellar yields in dedicated Galactic chemical evolution models for the Nuclear Stellar Disk.

A major observational advance is provided by the ALMA CMZ Exploration Survey (ACES), the largest ALMA mosaic ever performed \citep{longmore2026}. ACES maps the inner $\sim200$ pc of the Galactic Centre at $\sim1.5''$ resolution, combining the 12-m array, the Atacama Compact Array, and Total Power observations to recover structures from cloud scales down to dense cores. Together with complementary single-dish spectral surveys, ACES enables isotopic studies to extend from individual molecular clouds to the entire CMZ. Spatially resolved measurements of multiple isotopic systems will reveal variations driven by local chemistry and Galactic-scale processes, providing key constraints on gas accretion, chemical enrichment, and future Galactic chemical evolution models.
\vspace{-0.5em}
\section{Conclusions}

Isotopic ratios provide one of the most powerful tools available for investigating both the local chemistry and the large-scale evolution of the molecular gas in the Galactic center. The combination of sensitive molecular spectroscopy, observations of rare isotopologues, and dedicated chemical modelling allows physical conditions and nucleosynthetic signatures to be studied simultaneously.

The main conclusions of this work can be summarized as follows:

\begin{itemize}

\item Deuterium fractionation reveals the presence of a cold and relatively quiescent gas component embedded within the otherwise warm and turbulent environment of the CMZ. These narrow components likely represent the earliest stages of dense gas evolution and demonstrate that cold gas can survive even under the extreme conditions of the Galactic center.

\item New measurements of the carbon isotopic ratio based on HC$_3$N and HC$_5$N, including the first analysis using doubly substituted $^{13}$C isotopologues, yield $^{12}$C/$^{13}$C ratios significantly higher than the canonical value traditionally adopted for the Galactic center.

\item Astrochemical modelling indicates that isotopic fractionation alone cannot explain the observed carbon isotopic ratios. Instead, the measurements suggest that part of the molecular gas in the CMZ has undergone a different chemical evolution, consistent with scenarios involving gas inflow from the inner Galactic disk through the Galactic bar.

\item Future observations of additional isotopic systems, together with new Galactic chemical evolution models and large ALMA surveys such as ACES, will provide a comprehensive picture of the origin and evolution of molecular gas in the Galactic center.

\end{itemize}

The synergy between molecular spectroscopy, astrochemistry, and Galactic chemical evolution is opening a new window onto the history of the Galactic center. As increasingly sensitive observations become available, isotopic ratios will play a central role in connecting the chemistry of molecular clouds with the assembly history of the innermost regions of the Milky Way.
\vspace{-0.5em}
\section*{Acknowledgements}
The project that gave rise to these results received the support of a fellowship from the ”la Caixa” Foundation (ID 100010434). The fellowship code is LCF/BQ/PR25/12110012. L.C also acknowledges support from the grant PID2022-136814NB-I00 funded by the Spanish Ministry of Science, Innovation and Universities/State Agency of Research MICIU/AEI/10.13039/501100011033 and by ERDF, UE.
The author thanks O. Sipil\"a, D. Romano, J. Martín-Pintado, V. M. Rivilla, M. Sanz-Novo, I. Jim\'enez-Serra, S. Mart\'in, S. Zeng, and all collaborators whose work contributed to the results presented in this proceeding.
\vspace{-0.3em}
\bibliographystyle{aa}
\bibliography{bibliography2}

\end{document}